\documentclass[reprint,showpacs,preprintnumbers,amsmath,amssymb,superscriptaddress, aps, pra]{revtex4-1}

\eprint

\usepackage{graphicx}
\usepackage{dcolumn}
\usepackage{tabularx}
\usepackage{bm}
\usepackage{color}
\makeatletter
\def\Ddots{\mathinner{\mkern1mu\raise\p@	
\vbox{\kern7\p@\hbox{.}}\mkern2mu
\raise4\p@\hbox{.}\mkern2mu\raise7\p@\hbox{.}\mkern1mu}}
\makeatother
\usepackage[pdfborderstyle={/S/U/W 1}]{hyperref}

\usepackage{lineno}
\begin{document}

\title{Coherent electron trajectory control in graphene}

\author{Christian Heide} 
\email[E-mail: ]{christian.heide@fau.de}
\affiliation{Laser Physics, Department of Physics, Friedrich-Alexander-Universit\"at Erlangen-N\"urnberg (FAU), Staudtstrasse 1, D-91058 Erlangen, Germany}
\author{Takuya Higuchi}
\affiliation{Laser Physics, Department of Physics, Friedrich-Alexander-Universit\"at Erlangen-N\"urnberg (FAU), Staudtstrasse 1, D-91058 Erlangen, Germany}
\author{Heiko B. Weber}
\affiliation{Applied Physics, Department of Physics, Friedrich-Alexander-Universit\"at Erlangen-N\"urnberg (FAU), Staudtstrasse 7, D-91058 Erlangen, Germany}
\author{Peter Hommelhoff}
\email[E-mail: ]{peter.hommelhoff@fau.de}
\affiliation{Laser Physics, Department of Physics, Friedrich-Alexander-Universit\"at Erlangen-N\"urnberg (FAU), Staudtstrasse 1, D-91058 Erlangen, Germany}
\date{\today}

\begin{abstract}
	We investigate coherent electron dynamics in graphene, interacting with the electric field waveform of two orthogonally polarized, few-cycle laser pulses. Recently, we demonstrated that linearly polarized driving pulses lead to sub-optical-cycle Landau-Zener quantum path interference by virtue of the combination of intraband motion and interband transition [Higuchi \textit{et al}., Nature \textbf{550}, 224 (2017)]. Here we introduce a pulsed control laser beam, orthogonally polarized to the driving pulses, and observe the ensuing electron dynamics. The relative delay between the two pulses is a tuning parameter to control the electron trajectory, now in a complex fashion exploring the full two-dimensional reciprocal space in graphene. Depending on the relative phase, the electron trajectory in the reciprocal space can, for example, be deformed to suppress the quantum path interference resulting from the driving laser pulse. Intriguingly, this strong-field-based complex matter wave manipulation in a two-dimensional conductor is driven by a high repetition rate \textit{laser oscillator}, rendering unnecessary complex and expensive amplified laser systems. \\
	The corresponding Phys. Rev. Lett. is aviable at:\\ \url{https://doi.org/10.1103/PhysRevLett.121.207401}.
\end{abstract}%

\maketitle

Controlling electron trajectories in condensed matter systems by the field of light is a rapidly evolving area in ultrafast optics. For example, it has been demonstrated that the high harmonic generation process in solids is governed by the exact shape of the electron's trajectory, controlled with linearly \cite{Ghimire2011, Luu2015, Langer2016, Liu2017} and more recently elliptically polarized \cite{You2017, Saito2017, Yoshikawa2017} driving fields. Similarly, the generation of currents on the femtosecond timescale in dielectrics, allowing for current switching at terahertz to optical frequencies, has been demonstrated to be sensitive to the electron trajectory \cite{Schiffrin2013, Vampa2014, Ndabashimiye2016, Chen2018, Schubert2014, Hohenleutner2015}. This coherent control opens the door for investigating electronic and topological properties in solid state systems at optical frequencies \cite{Vampa2015, Kelardeh2016, Tancogne-Dejean2017, Tancogne-Dejean2017_1}.

In contrast, much less is known about coherent electron trajectory control in metallic or narrow bandgap materials. In the case of metals, charge carriers screen an external electric field, consequently it is difficult to apply strong electric fields to metals \cite{JohnDavidJackson1999, Schiffrin2013, Krausz2014}. In semiconductors, resonant absorption results in heating and damage of the material when illuminating the sample with high laser intensities. To overcome these difficulties, we use graphene. Even though the metallic nature of graphene is reflected in its excellent carrier mobility \cite{Novoselov2012}, the carrier concentration is low compared with conventional metals and thus screening due to free carriers is negligible at optical frequencies. Therefore, strong optical fields can be generated in graphene. In addition, graphene, in particular epitaxial graphene on SiC(0001), is one of the most robust materials available \cite{Emtsev2009}, and can withstand high laser intensities. This implies that graphene represents an ideal material to study light-field-driven dynamics in conducting materials.

\begin{figure*}[t]
	\begin{center}
		\includegraphics[width=16cm]{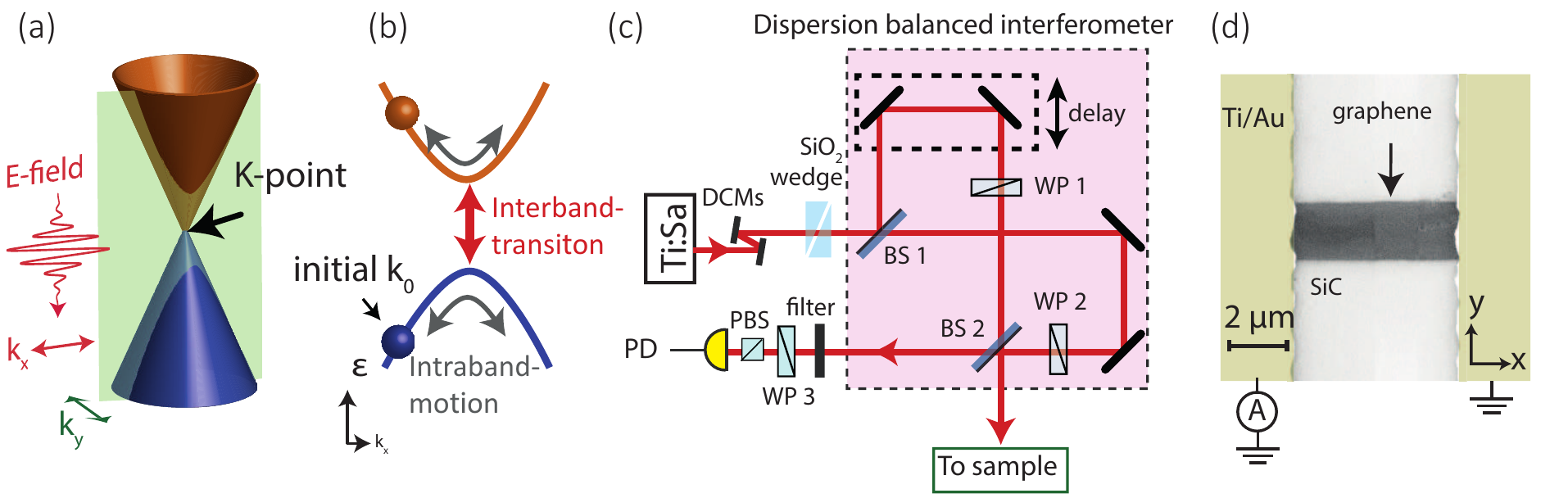}
		\caption{\label{Figure1} (a) Electronic band structure of graphene around the K-point. The green area indicates a condition where both interband transition and intraband motion are present. (b) Combined interband and intraband dynamics. (c) Experimental setup; CEP-stabilized laser pulses from a Ti:Sa laser are pre-chirped using double chirped mirrors (DCMs) and SiO$_2$ wedges to achieve the shortest laser pulses at the sample. The laser beam is sent into a Michelson interferometer to split it into two beams with a beamsplitter (BS 1), rotate the polarization of the control pulses by 90$^\circ$ with respect to the driving pulses’ polarization using a half waveplate (WP 1) and combine them afterward (BS 2). The temporal delay is modulated with a piezo element and can be measured using optical interference after propagating through WP 3, a polarizing beam splitter (PBS) and a narrow-bandpass filter. A photodiode (PD) records the optical interference. (d) Scanning electron micrograph of the graphene sample. Graphene, epitaxially grown on SiC, is patterned to obtain a graphene stripe (2\,$\mu$m $\times$ 5\,$\mu$m), contacted with two titanium/gold electrodes. In the experiment the electrodes are not illuminated ($I_\text{Ti/Au}$/$I_\text{peak}<$ 2$\times$10$^\text{-5}$, with $I_\text{Ti/Au}$ the intensity at the electrodes and $I_\text{peak}$ the intensity in the center of the graphene stripe). The sample is placed in a vacuum chamber with a base pressure of 1$\times$10$^{-7}$\,mbar at room temperature.}
	\end{center}
\end{figure*}

We have recently shown that few-cycle laser pulses generate a residual current sensitive to the waveform of the driving laser pulses. This current scales non-monotonically as a function of the field strength \cite{Higuchi2017} and results from breaking the spatiotemporal symmetry of the system by the few-cycle nature of the laser pulses \cite{Franco2008}. When the light field is strong enough so that the light-field-driven change of the electron wavenumber during the interaction dominates the overall light-matter interaction, the resultant current direction can be predominantly determined by electron trajectories in the reciprocal space. 
In this regime the temporal evolution of the electron wavenumber is proportional to the electric field strength $\dot{\textbf{k}}(t) \propto \textbf{E}(t)$ \cite{Kittel1963, Golde2008}. 
In graphene this light-field-driven dynamics can be reached with a moderate field strength of 2\,V/nm. Note that due to graphene's particular band structure the critical field strength required to enter the strong-field regime is about one order of magnitude smaller than that for gas-phase atoms \cite{Krause1992} or dielectrics \cite{Ghimire2011, Schiffrin2013}. 

In this strong-field regime, free electrons in graphene are driven by the light field. These electrons follow trajectories nearby an avoided crossing formed by the conduction and the valence band. The dynamics around such a crossing can be described by the Landau-Zener (LZ) transition framework \cite{Zener696, Shevchenko2010}. When the driven electron enters the avoided crossing region, its wavefunction can be coherently split: One part of the wavefunction can stay in its original band (intraband motion), whereas the other part can experience an interband transition (Fig.~\ref{Figure1}(b)). Thus, this action represents a beamsplitter for the driven electrons in the reciprocal space. Since the driving waveform is an oscillating electric field, this beamsplitter action occurs twice within one optical cycle. When the driving optical period is shorter than any electronic dephasing time, the electronic coherence between two LZ events is maintained and the different quantum-pathways can lead to interference effects, known as Landau-Zener-St\"uckelberg (LZS) interference \cite{Shevchenko2010}. More precisely, for linearly polarized driving fields, electrons are driven back and forth along a one-dimensional trajectory, allowing for the mentioned two Landau-Zener transition events (Figs.~\ref{Figure1}(a),~\ref{Figure1}(b)). For instance, the electron can first undergo an intraband motion followed by an interband transition or vice versa. Depending on the phase evolution of each quantum path the interference can be constructive or destructive resulting in an excitation or no excitation. As a consequence of this interference the resulting current direction can be controlled by the exact shape of the waveform. 

In contrast, by employing circular polarization the electron trajectory is two-dimensional, suppressing intra-optical-cycle interference. In this case, the transition probability is rather determined by the distance in \textit{k}-space of the electron trajectory to the Dirac point, which defines the magnitude of the optical transition matrix element. This distance strongly depends on the waveform of the laser pulse, which is why also for circular polarization a waveform dependence is observed.\\

In this letter, by employing laser pulses with various degrees of ellipticity, we investigate how the nature of the two waveform-dependent current generation processes transition from one to the other. Because the current generation is qualitatively different in the two extreme cases (linear vs. circular polarization), as discussed in the preceding paragraph, this transition is nontrivial. More generally, we investigate complex, i.e. full 2-d, strong-field-based electron matter wave control.

In the experiment we focus two few-cycle pulsed laser beams at a graphene stripe (Figs.~\ref{Figure1}(c),~\ref{Figure1}(d)). One pulsed beam is polarized linearly and parallel to the stripe direction, leading to a current based on LZS interference \cite{Higuchi2017}. We call these laser pulses the driving laser pulses. The other pulsed beam, comprising of the control pulses, is also linearly polarized, but perpendicularly to the graphene stripe direction. These pulses alone do not generate a measurable waveform-dependent current because they do not break the mirror symmetry along the current measurement direction. The two pulsed beams can be delayed in time with respect to each other, allowing us to generate laser pulses with arbitrary ellipticity.

To generate the two independently controlled pulsed beams, we steer carrier-envelope phase- (CEP)-stable laser pulses from a Ti:Sa oscillator with a repetition rate of 80\,MHz and a Fourier-limited pulse duration of 5.5\,fs (FWHM) into a Michelson interferometer (Fig.~\ref{Figure1}(c)), in which the polarization in one arm is rotated by 90$^\circ$. The resulting beam with the two pulses is focused to a spot with 1.5\,$\mu$m radius at the graphene sample (Fig.~\ref{Figure1}(d)). To pick up the current sensitive to the pulse waveform we modulate the CEP by setting the carrier-envelope-offset frequency $f_{\text{CEO}}$ to 1.1\,kHz. With a lock-in measurement scheme and $f_\text{CEO}$ as reference the CEP sensitive current is recorded.

\begin{figure}[t]
	\begin{center}
		\includegraphics[width=7cm]{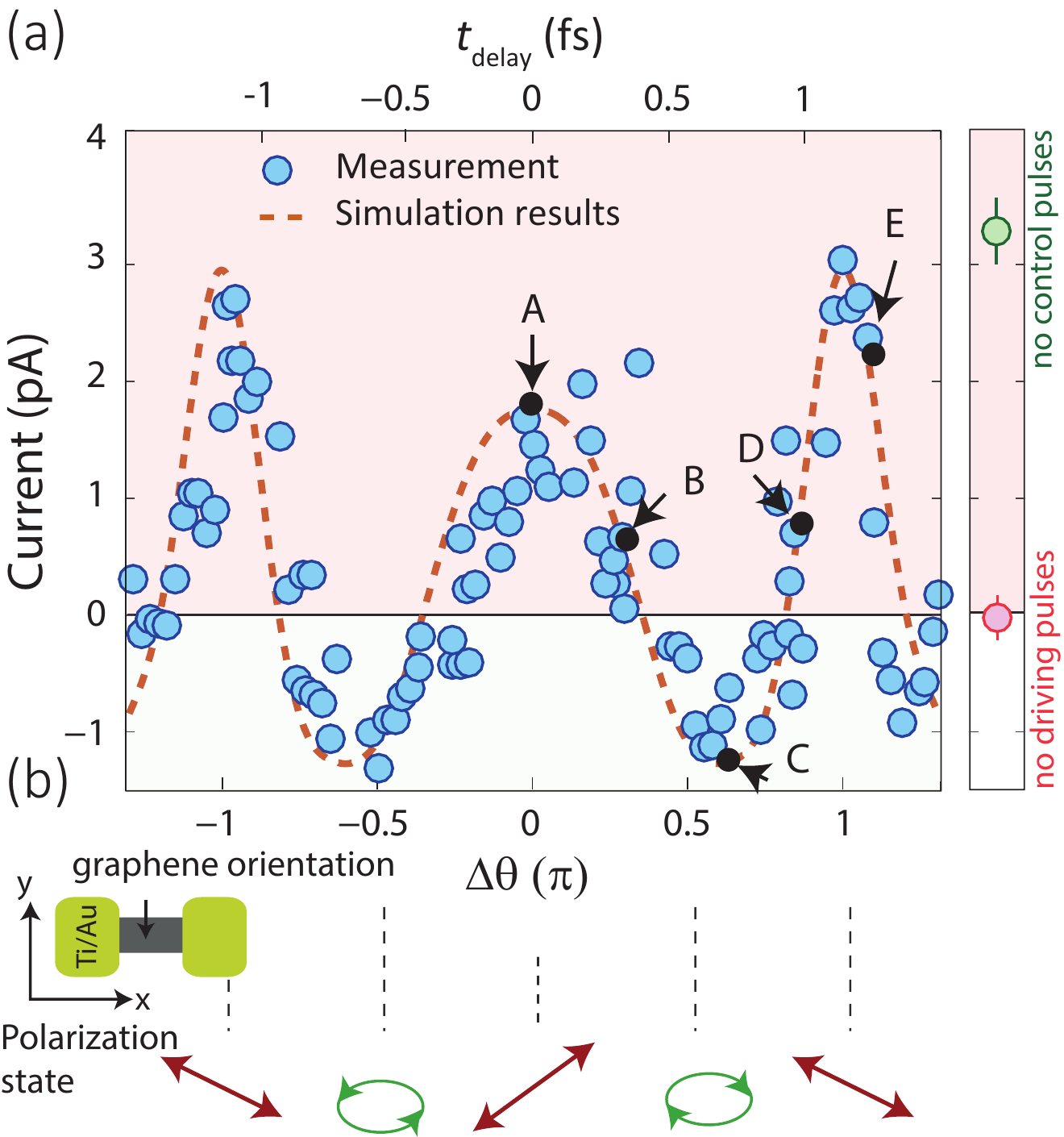}
		\caption{\label{Figure2} (a) Measured CEP-dependent current as a function of $t_\text{delay}$ between two orthogonally polarized pulsed laser beams. The difference between experimentally measured currents generated by pulses with $\Phi_\text{CEP}=-\pi$/2 and $\Phi_\text{CEP}=\pi$/2 is plotted (blue dots). In the red-colored area the current is positive, indicating the LZS regime, whereas in the green-colored area the current is negative and LZS is turned off. The simulation result is plotted as orange dashed line. The markers A-E refer to Fig.~\ref{Figure3}. Clearly, the simulation reproduces the data very well, including the smaller peak current at $\Delta\theta=0$ as compared to $\Delta\theta=\pm \pi$. The currents for these values differ because of the shortness of the laser pulse. When the control pulses is blocked the current amplitude is positive, whereas for a blocked driving pulses no measurable CEP-dependent current is recorded (right panel).
			(b) The respective main polarization state (here simplified for a single optical cycle) is depicted for different delays between the control and driving pulse. Given that the driving pulses (E$_\text{drive}$ =2.2\,V/nm) are polarized parallel to the graphene stripe ($x$ direction) and the control pulses (E$_\text{control}$ =1.6\,V/nm) perpendicular to it, the resulting polarization is rotated by 36$^\circ$ relative to the \textit{x} axis in the case of 0 delay.}
	\end{center}
\end{figure}

Fig.~\ref{Figure2}(a) shows the measured CEP-dependent current as a function of the temporal delay between the control and the driving pulsed laser beam. The peak field strength of the driving pulses is $E_{\text{drive}} = 2.2$\,V/nm at the graphene sample (to reach the strong-field regime), while the control pulses' field strength is $E_{\text{control}} = 1.6$\,V/nm. Strikingly, we observe a reversal of the current direction when we change the delay between the driving and the control pulses. The polarization state as a function of the delay is depicted in Fig.~\ref{Figure2}(b). When the delay is 0, resulting in linear but rotated polarization, a positive current with $J_\text{linear} = 1.8\pm 0.2$\,pA is observed. In contrast, for a temporal delay of $\pm 0.6$\,fs, corresponding to elliptic polarization, a negative current with $J_\text{elliptic} = -1.2 \pm 0.2$\,pA is found. Increasing the temporal delay to $\pm 1.1$\,fs results once again in a linear polarization state and a positive current amplitude of $3.0 \pm 0.2$\,pA.\\

To understand the change in current direction as a function of the delay between the control and driving pulses, we compare the experimental observation with numerical simulation results \cite{Neto2007, Higuchi2017}. The conduction band population after excitation with two laser pulses is modeled by numerically integrating the Schr\"odinger equation for electrons in graphene. 
The time-dependent Hamiltonian describing the electron dynamics around the $K$-point of graphene can be well described by the Dirac-Weyl Hamiltonian:
\begin{equation}
{\cal H}(t) = v_{\rm F} \boldsymbol{\sigma} \cdot {\bf p}(t),
\end{equation}
where $\boldsymbol{\sigma}$ are the Pauli matrices and the momentum ${\bf p}(t) = \left(p_x(t), p_y(t), p_z(t) \right)$ evolves in time based on the Peierls substitution ${\bf p}(t) \equiv {\bf p}_0(t) - e \textbf{A}(t)$ \cite{Ishikawa2010}.
The waveform of a single pulse is chosen such that the envelope of the vector potential $\textbf{A}(t)$ is a Gaussian function with a pulse duration of 5.5\,fs (FWHM) and a central frequency of 375\,THz. The peak electric field values are 2.4\,V/nm and 1.8\,V/nm for the driving and the control pulses, respectively. These waveforms are chosen because they closely reproduce the main part of the experimental optical waveforms, whereas they are simplified to keep the computational effort manageable. A detailed description of the simulation can be found in \cite{Higuchi2017}.

\begin{figure*}
	\begin{center}
		\includegraphics[width=15cm]{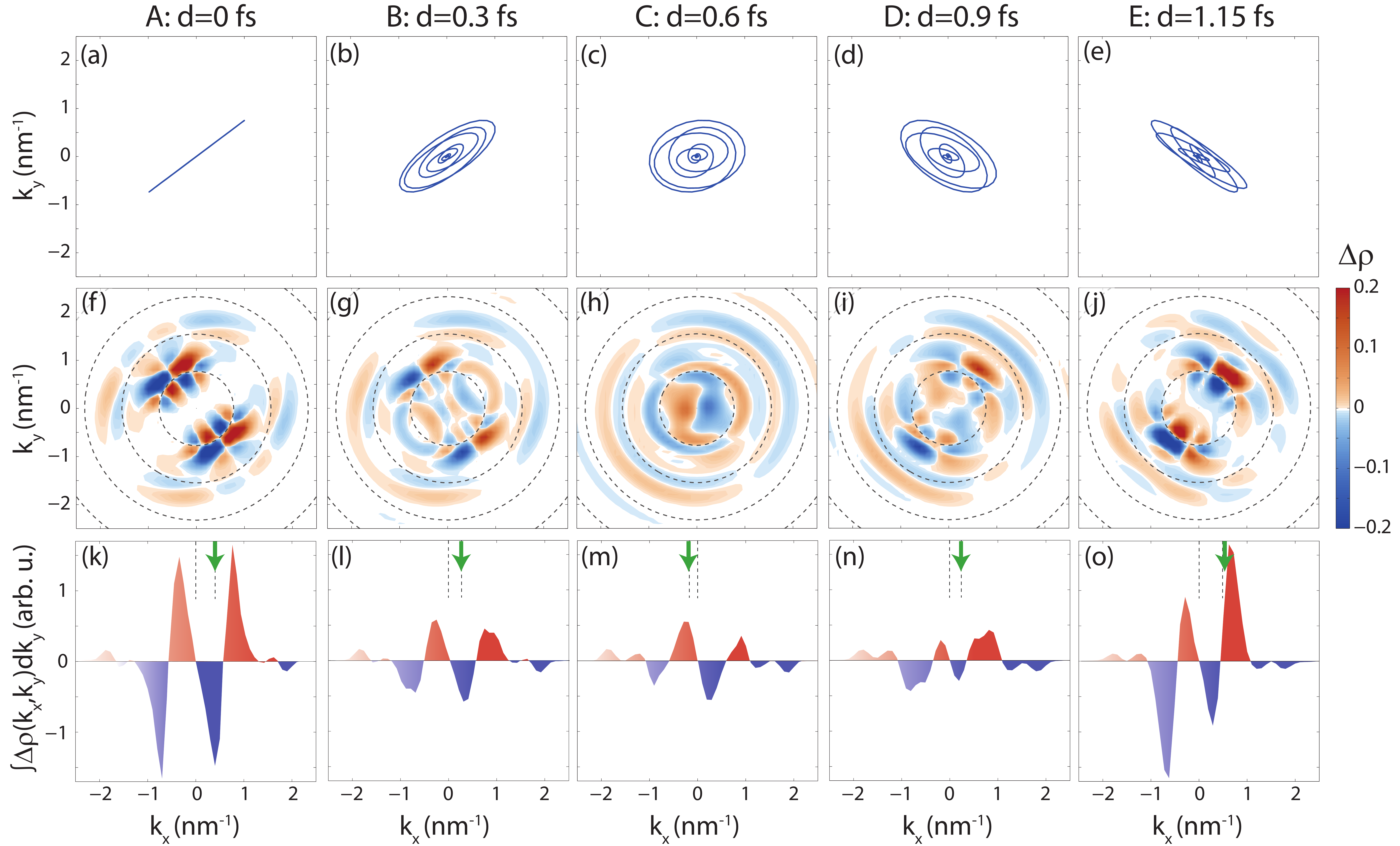}
		\caption{\label{Figure3}(a) to (e) Electron trajectories in reciprocal space, plotted for various delays between the driving and the control pulses. The five panels correspond to the delay values marked in Fig.~\ref{Figure2} (A to E). The initial wavenumber is $k_x=k_y=0$. Clearly, the electrons sample k-space decisively different, from a line to a spiral-out-spiral-in movement.
			(f) to (j) CEP-dependent electronic current density $\Delta\rho = \rho_\text{C}(\Phi_\text{CEP}=\pi/2)-\rho_\text{C}(\Phi_\text{CEP}=-\pi/2)$. The red areas indicate that pulses with $\Phi_\text{CEP}=\pi/2$ generate more excitation than ones with $\Phi_\text{CEP}=-\pi/2$. For \textit{t}$_\text{delay}$ = 0 or 1.15\,fs, the red areas can be found more at $k_x > 0$. In contrast, for \textit{t}$_\text{delay}$ = 0.6\,fs at $k_x < 0$, indicating a current reversal. 
			The dotted circles indicate energies corresponding to (multi-) photon resonances. The energy difference between conduction and valence band corresponds to $\hbar \omega$ on the innermost circle, and the subsequent rings correspond to $2\hbar \omega$ and $3\hbar \omega$. (k) - (o) Electronic current density integrated along $k_y$. The center of mass of the CEP-dependent conduction band population is indicated by the green arrow. The resulting current is plotted in Fig.~\ref{Figure2} as dashed line.}
	\end{center}
\end{figure*}

Figs.~\ref{Figure3}(a)~-~\ref{Figure3}(e) show the light-field-driven electron trajectories plotted for different delay values $t_\text{delay}$ between the driving and the control pulses and ${\bf p}(t=0)=0$, resulting from the model. We calculate the conduction band population after excitation with these pulses. To highlight the CEP-dependent excitation, we show in Figs.~\ref{Figure3}(f)~-~\ref{Figure3}(j) the difference $\Delta\rho_{\rm C}({\bf p}_0)$ in conduction band population between excitations with $\Phi_\text{CEP}=\pi/2$ and $\Phi_\text{CEP}=-\pi/2$, calculated for different delay values. We chose these particular $\Phi_\text{CEP}$ values because they yield the largest contribution in the LZS case \cite{Higuchi2017}.

When the light is (almost) linearly polarized, i.e., $t_\text{delay}=0$ (Fig.~\ref{Figure3}(a)) or $1.15$\,fs (Fig.~\ref{Figure3}(e)), $\Delta\rho_{\rm C}({\bf p}_0)$ shows an (almost) anti-symmetric distribution.
In these two cases, we find red areas more at $k_x>$0, where the group velocity $v_x$ along the $x$ direction is positive. The CEP-dependent electronic current density (i.e., difference between $\Phi_\text{CEP}=\pi/2$ and $\Phi_\text{CEP}=-\pi/2$) along the $x$ axis after this excitation can be described as
\begin{equation}
j_\text{CEP} = \int dp_x dp_y v_x({\bf p}_0) \Delta\rho_{\rm C}({\bf p}_0),
\end{equation}
which takes a positive value for $t_\text{delay}=0$ and $t_\text{delay}=1.15$\,fs. Together with the negative sign of $e$, the current flows to the negative direction for $\Phi_\text{CEP}=\pi/2$ and to the positive direction for $\Phi_\text{CEP}$ = $-\pi/2$.

When the driving and the control pulses are overlapped with a delay of $t_\text{delay}=0.6$\,fs, the electron trajectory is almost circular. Therefore, the electrons do not pass the same point twice within an optical cycle, suppressing LZS interference in \textit{k}-space \cite{Higuchi2017}. Indeed, the red areas in Fig.~\ref{Figure3}(h) are now found rather at $k_x<$0, which leads to the generation of a current with opposite sign compared to the case of $t_\text{delay}=0$\,fs. To highlight the difference in the current density we integrate $\Delta \rho(\textbf{k})$ along $k_y$. The center of the CEP-dependent conduction band population is shifted along $k_x$ (Figs.~\ref{Figure3}(k)-(o)), resulting in a positive or negative current.
These results demonstrate that the sign of the current can be controlled with a second, weaker control pulse, by adjusting the relative delay between the driving and the control pulse, and hence the \textit{k}-space trajectory. 

To compare the simulation with the experiment, we calculate the residual current from the conduction band population. We assume a ballistic carrier lifetime of 40\,fs and a diffusive decay length of 350\,nm, consistent with previous literature \cite{Breusing2011, Gierz2013}. The result of the simulation is plotted in Fig.~\ref{Figure2} and shows excellent quantitative agreement with the experimental data. For this reason, we trust the simulation to yield proper electron trajectories in \textit{k}-space. 

Combining this electron trajectory control with the well-established method of polarization gating \cite{Corkum1994, Tcherbakoff2003, Villoresi2006}, for example, would even allow us to generate laser pulses with time-varying ellipticity to control the LZS interference at optical carrier frequencies. In such a polarization-gated pulse, the polarization varies from circular to linear and then back to circular, providing a window of linear polarization which can be as short as a single optical cycle \cite{Shan2005}. This way, the interference phenomena can be constrained in time to (less than) one optical cycle.\\

In summary, we extend complex two-dimensional coherent matter-wave control to an important and new material class - 2D materials and conductors. The temporal delay between two orthogonally polarized laser fields is a tuning knob to tailor the ellipticity of the driving laser pulses which enables us to control intra-cycle LZS interference, i.e., turn it on or off. The resulting change in current direction as a function of the temporal delay between the two laser pulses indicates that the electron trajectories in the reciprocal space are deformed, suppressing the matter-wave interference.
These electron dynamics in graphene takes place on timescales faster than electron-electron and electron-phonon scattering, which is why the coherent control of electrons inside of graphene works so well, namely without noticeable dephasing, fully coherent. 
It will be interesting to explore the question: What are the dominating mechanisms and timescales for dephasing? The required comparably small field strength to study light-field-driven phenomena in solid state systems can be achieved with a commercial laser oscillator, without requiring large and complex laser amplifier systems, which may promote a widespread use of this technique.
In the future, we envision not only the extraction of (de-)coherence time scales of the electron wavefunction in solids but also 2-dimensional band structure tomography. The LZS physics is not limited to graphene but will likely be found in other material systems as well. Furthermore, if the light field of the control pulses contains only a single optical cycle, the influence of the control pulses can be used to sequentially scan the cycles of the driving pulses, a technique that could be used for light field retrieval.\\

This work has been supported in part by the European Research Council (Consolidator Grant NearFieldAtto) and the Deutsche Forschungsgemeinschaft (grant SFB 953).
\end{document}